\documentclass{PoS}
\usepackage{enumitem}
\title{Prospects for Gamma-Ray Bursts detection by the Cherenkov Telescope Array}
\ShortTitle{Prospects for GRB detection by CTA}
\author{
\speaker{E.~Bissaldi}$^{\; (1)\;}$,  
T.~Di Girolamo$^{\;(2,3)\;}$,
F.~Longo$^{\;(4,5)\;}$, 
P.~Vallania$^{\;(6,7)\;}$, 
C.~Vigorito$^{\;(7,8)\;}$
$\;\;\;\;$for the CTA Consortium\footnote{Full consortium author list at {\tt http://cta-observatory.org}} \\
$^{(1)}$ Istituto Nazionale di Fisica Nucleare, Sezione di Bari, 
Bari, Italy \\
$^{(2)}$ Dipartimento di Fisica, Universit\`a degli Studi di Napoli, Napoli, Italy \\
$^{(3)}$ Istituto Nazionale di Fisica Nucleare, Sezione di Napoli, 
Napoli, Italy  \\
$^{(4)}$ Dipartimento di Fisica, Universit\`a degli Studi di Trieste, 
Trieste, Italy \\
$^{(5)}$ Istituto Nazionale di Fisica Nucleare, Sezione di Trieste-Udine,
Trieste, Italy \\
$^{(6)}$ Osservatorio Astrofisico di Torino dell'Istituto Nazionale di Astrofisica, 
Torino, Italy \\
$^{(7)}$ Istituto Nazionale di Fisica Nucleare, Sezione di Torino, 
Torino, Italy \\
$^{(8)}$ Dipartimento di Fisica, Universit\`a degli Studi di Torino, 
Torino, Italy
\\
{\footnotesize{E-mail:}} {\tt{\footnotesize{Elisabetta.Bissaldi@ba.infn.it}}}}
\abstract{
The first Gamma--Ray Burst (GRB) catalog presented by the Fermi-Large Area Telescope (LAT) collaboration includes 28 GRBs, detected above 100 MeV over the first three years since the launch of the Fermi mission. However, more than 100 GRBs are expected to be found over a period of six years of data collection thanks to a new detection algorithm and to the development of a new LAT event reconstruction, the so-called ``Pass 8''. Our aim is to provide revised prospects for GRB alerts in the CTA era in light of these new LAT discoveries. \\
We focus initially on the possibility of GRB detection with the Large Size Telescopes (LSTs). Moreover, we investigate the contribution of the Middle Size Telescopes (MSTs), which are crucial for the search of larger areas on short post trigger timescales. The study of different spectral components in the prompt and afterglow phase, and the limits on the Extragalactic background light are highlighted. Different strategies to repoint part of -- or the entire array -- are studied in detail.}
\FullConference{The 34th International Cosmic Ray Conference,\\
		30 July- 6 August, 2015\\
		The Hague, The Netherlands}
\begin{document}
\section{Introduction}
Gamma--Ray Bursts represent a very interesting
case study in astrophysics, mainly due to their
multi--disciplinary nature. 
At present time, a GRB can trigger one or more of the 
dedicated instruments based on several satellites
orbiting around the Earth, such as Swift, 
Fermi, MAXI or INTEGRAL.
The observed keV-MeV prompt emission may be
accompanied by X--ray, optical or radio emission.
Rapid follow--up of this keV--MeV emission is possible thanks to
communication through the Gamma-ray Coordinates Network (GCN), where the
GRB position is spread out in real time to all other
observatories. This includes all currently
operative Imaging Atmospheric Cherenkov Telescopes (IACTs) like
MAGIC, H.E.S.S., and VERITAS.
Unfortunately, none of them ever succeeded in capturing
a high-energy signal from a GRB, but several upper limits
from a single or from a sample of bursts
were published by each collaboration over
the last years.
In this contribution, we aim to investigate 
the possibility by the 
Cherenkov Telescope Array (CTA) \cite{CTA}
to detect such elusive GRB emission.
\section{CTA configurations}
CTA is a worldwide 
project aiming to build and operate a third generation of IACTs.
In its current design, two huge arrays for a 
total amount of more than 100 telescopes are 
foreseen, one in each hemisphere, to extend 
the energy range of currently operating IACTs
mainly to higher energies and to improve the 
sensitivity of about one order of magnitude 
with better angular and energy resolutions.

In order to cover the energy range from 20 GeV 
to more than 100 TeV, three kinds of telescopes are envisaged: 
Large Size Telescopes (LSTs, 20 GeV $\div$ 100 GeV), 
Medium Size Telescopes (MSTs, 100 GeV $\div$ 10 TeV), 
and Small Size Telescopes (SSTs, 10 TeV $\div$ > 100 TeV).
This configuration is driven by the features 
of the Cherenkov signal at different energies: 
near the threshold, the number of source photons 
is relatively high but the Cherenkov image 
is poor, so few huge telescopes 
(3 or 4, with a diameter of 23 m) 
are used to collect the faint showers.
In this region, the challenge is to distinguish 
between Cherenkov and Night Sky Background (NSB) 
photons, and to discriminate primary gammas 
from the overwhelming flux of cosmic rays hadrons.
On the other hand, the effective area is not 
an issue, so a small number of big telescopes 
close to each other is the best configuration.
At higher energies, the Cherenkov signal starts 
to increase and the flux from the source is 
rapidly fading, so a compromise array of 
$\sim$ 25 telescopes of 12 m diameter scattered 
over an area of around 3 km$^2$ is the best choice.
At the highest energies the situation is 
completely different: the Cherenkov signal 
is very strong, but the steep spectra reduce 
the signal to a handful of primary gammas.
To overcome this, a huge effective area up 
to 6-7 km$^2$, covered by $\sim$ 70 SSTs (4 m diameter) is needed.
We notice that for energies $\gtrsim$ 10 TeV the 
Cherenkov images are so rich that it is not only 
easy to discard the NSB photons, but also to 
discriminate gammas from hadrons, 
approaching a background-free working 
mode for which the sensitivity is proportional 
to the effective area and not to its square root.
Due to the Extragalactic Background Light (EBL) absorption, 
that strongly depresses the high--energy spectra 
of distant sources detected on Earth 
(the limit being E $\lesssim$ 1 TeV 
for z $\geq$ 0.1) and to the fact that 
the Galactic plane is mostly visible 
from the Southern hemisphere, the SST array 
is not expected to be built in the 
Northern observatory.
\section{High--energy GRB observations}
Due to their elusive nature, GRBs represent
a very interesting candidate for future observations by
CTA, as highlighted also in \cite{Susumu}. 
Prospects for VHE GRB observations by CTA
were also presented in detail 
by \cite{Gilm}.
Most of these analyses relied on extrapolations taken
(1) from the GRB spectral parameters published 
in the catalogs of the Burst and
Transient Source Experiment (BATSE,
20 keV--2 MeV) or of the {\it Swift}
Burst Alert Telescope (BAT,
15--150 keV); and (2)
from some very energetic GRBs detected
by Fermi before 2012.

Since June 2008, the Fermi mission 
is constantly enhancing the number of GRB 
detections.
The Fermi Gamma-Ray Burst Monitor (GBM, \cite{Meegan})
is sensitive in the energy rate bewteen 
8 keV to 40 MeV and its field of view
covers almost 4 $\pi$ sr. The GBM trigger rate
lies at $\sim$250 GRB/yr (i.e. higher than
BAT with $\sim$100 GRB/yr) and the second GBM catalog
(covering 4 years of operation) includes almost 
1000 bursts \cite{AvK2}.
Furthermore, thanks to the LAT
\cite{Atwood} onboard Fermi, the number of high--energy GRBs
dramatically increased with respect to the bunch of events
seen in the '90s by the Energetic Gamma--ray Experiment Telescope 
(EGRET, 20 MeV--30 GeV).
LAT operates at energies between 100 MeV and $>$300 GeV
and its first GRB catalog \cite{AckGRB} already included
$\sim$30 GRBs observed during the 
first three years of operation
above 100 MeV with the standard event reconstruction
analysis. This number gets even larger 
when considering dedicated
low--energy techniques (LLE) exploring the energy
region between 10 and 100 MeV, where the LAT and GBM
energy ranges overlap.

The current number of official LAT GRB
detections is constantly kept up--to--date on the 
LAT Public Table website \cite{GRBtable} and lists
almost 100 GRBs at the time of this writing,
that is after seven years into the Fermi mission.
However, this number is meant to grow as soon
as the new event reconstruction algorithm
(the so--called {\it Pass 8}) will be released
and the catalog will be updated (see \cite{Giacomo}
for more details).

Fermi's GRBs exhibit dozen of GeV photons 
with unprecedented high energies.
The current record holder is GRB 130427A,
which emitted a 95 GeV photon \cite{LAT-130427A} and was
detected at a low redshift of 0.34. It was followed--up
by a very large number of telescopes, including
the Veritas array \cite{VERITAS-130427A}. 
Unfortunately, Veritas' observations only 
began almost 20 hours post
trigger, leading to no GRB detection. The upper limit (UL)
was placed at 3.3 $\times\,10^{\,-12}$ erg
cm$^{\,-2}$ s$^{\,-1}$.

In the near future, we plan to revisit the latest prospects
for VHE GRB observations in the light of the
newest simulations produced by the CTA
collaboration.
We want to focus on the extrapolation to
high energies of Fermi--like GRBs, both 
from the prompt emission using the GBM sample
and from the late emission using the LAT sample,
in particular using those bursts with measured redshift.
For this contribution, we use 
the extremely fluent GRB 130427A
as a test case and see how well we might observe
this event at several epochs after the trigger 
with various CTA configurations.
\section{Simulation of GRB observations with ctools}
In order to estimate the possible detection of GRBs 
by CTA, we intend to set up a library of GRBs at 
different times, extrapolating the LAT flux 
to the highest energies and properly taking into 
account the time evolution of their flux. 
However, since we do not yet model the effect of 
the EBL, our current simulations are limited up to 100 GeV,
an energy where the effect of the cosmological attenuation 
is not relevant (see Section 5. for more details). 

To estimate the detectability of a GRB we make use of 
the {\it ctools}, a software package 
specifically developed for 
the scientific analysis of CTA data \cite{ctools}. 
In these first tests, we simulate the high--energy 
emission of GRB 130427A as detected by LAT, 
with a spectral index $\gamma = 2.2$ 
(which is almost constant from 400 s
up to 70 ks post trigger)
and a power law decay
with a temporal index $\tau = 1.35$
(valid for t $>$ 380 s).
We cross checked our extrapolation
to the CTA energies by comparing it with the UL placed
on this GRB by VERITAS \cite{VERITAS-130427A}.   
This was calculated assuming an intrinsic GRB spectrum of 
(dN/dE) $\propto$ E$^{−\gamma}$  
absorbed using the EBL model of \cite{Gilmore}.

\begin{figure}[t!]
\begin{tabular}{cc}
\centering
\includegraphics[height=0.23\textheight,bb=0 0 680 494]{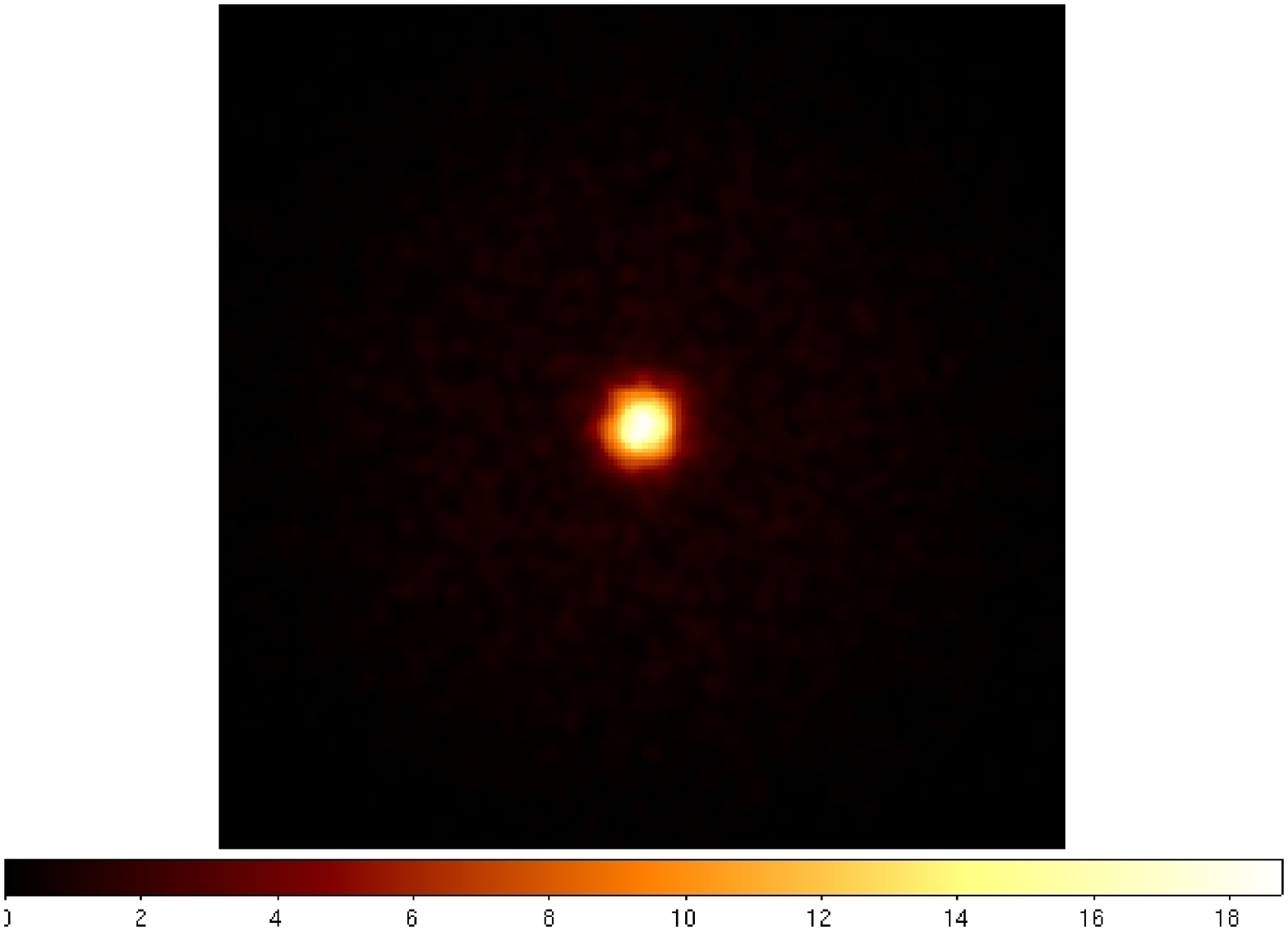} & 
\includegraphics[height=0.23\textheight,bb=0 0 680 494]{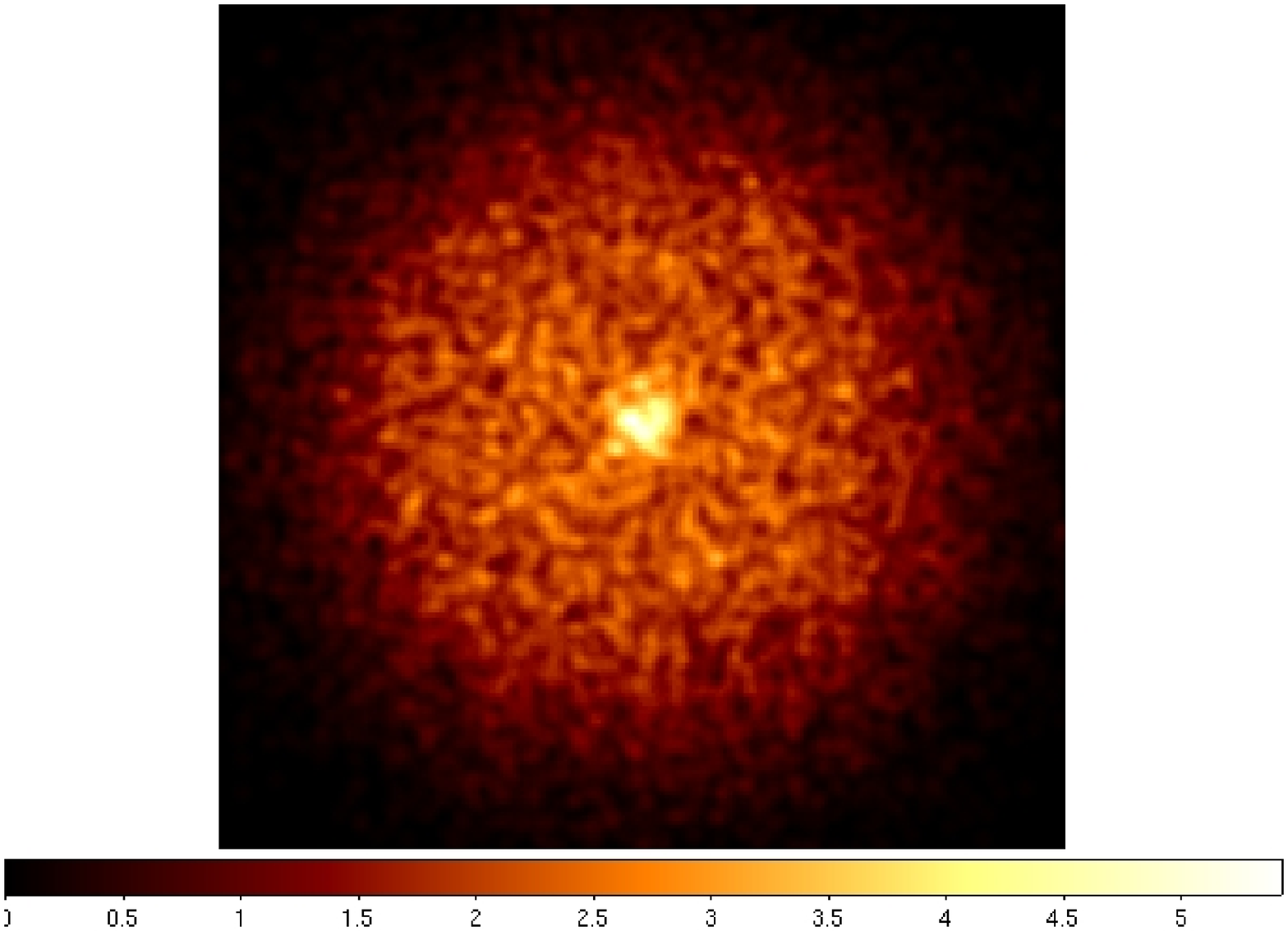}
\end{tabular}
\caption{Simulation of GRB 130427A with {\it ctools}. 
{\it Left panel}: 
Count map of GRB 130427A as seen by 
CTA North in 600 s, after 1000 s from the trigger 
in the energy range 20 GeV--100 GeV; 
{\it Right panel:}
Count map of GRB 130427A as seen by CTA North in 1800 s, 
after 10000 s of the trigger in the energy range 
20 GeV--100 GeV}
\label{fig:grb130427A}
\end{figure}

We simulate two possible observations of this GRB by CTA: 
(a) the first one lasting 10 min
at t $=$ 1 ks post trigger; (b) the second one 
lasting half an hour at t $=$ 10 ks post trigger. 
We assume both observations on axis with respect 
to the CTA array and a zenith angle $\theta = 20^{\,\rm o}$.
Figure \ref{fig:grb130427A} shows two
count maps of the simulated GRB in the 
energy range 20 GeV--100 GeV.
They were obtained using
the {\it ctools} functions {\it ctobssim} and {\it ctbin}, 
and adopting the recently provided CTA
instrument response functions (IRFs)
\cite{response-functions}. 
In this particular case, we made use of 
the {\it North\_0.5h} and {\it North\_5h} IRFs, 
respectively. A preliminary {\it ctlike} 
analysis is performed on the two observations, 
getting a significant detection in both cases. 
We plan to perform similar analyses
for other Fermi GRBs with known redshift, 
using different observing 
profiles and spectra. 
\section{Effect of the Extragalactic Background Light}
The interaction of extragalactic VHE photons 
with EBL at UV--optical wavelengths produces an 
$e^+ e^-$ pair and thus an exponential attenuation 
of the gamma--ray flux.
This absorption increases with the redshift 
of the source, the gamma--ray energy
and the photon density of the EBL 
(for a review, see \cite{Dwek}). Moreover,
it is particularly important for GRBs, 
since they are cosmological sources with a 
mean redshift $z \simeq 2$.
Many EBL models have been published in 
the last decades, with a general
trend towards a decrease of the 
corresponding optical depth due 
to the observation of VHE 
photons at larger redshifts 
\cite{Acciari}, which were 
recently detected from blazars 
at $z \simeq 1$ \cite{Mir1,Mir2,Muk}. 
The EBL models are getting close to the 
firm lower limits derived from 
integrated galaxy number counts \cite{Madau}. 
As an indication, the high--energy spectrum 
of a source at z$\simeq1$ shows a 
cutoff at E$\approx$100 GeV.
In our future work we will consider the 
EBL model by \cite{Franc}.
For simplicity, in our simulations we will extend 
the spectrum only up to 1 TeV, 
the maximum energy after which the source 
is assumed to be totally absorbed.

However, the observed spectral indices in 
blazars do not seem to follow the 
amount of softening with increasing 
redshift predicted by the EBL absorption 
\cite{DeAng}, and there are other indications 
of an overestimation of the EBL photon density 
(e.g., \cite{Rubt}). A possible explanation of these 
results could be the oscillation from 
photons to Axion--Like Particles (ALPs), 
which propagate unimpeded over cosmological 
distances before reconversion, 
reducing the optical depth along the VHE gamma--ray path 
\cite{DeAng11}.
In this framework, GRBs, with their cosmological 
distances, may be useful to add
stronger constraints on the EBL and give 
new hints on the existence of ALPs.
\section{CTA operating modes}
CTA's complex and varied experimental 
layout will be run in different ways 
depending on the target features:
\begin{enumerate}[topsep=1pt, partopsep=1pt,itemsep=1pt,parsep=0pt]
   \item{highest sensitivity observations, with all telescopes pointing toward a single source;}
   \item{normal operations, with the array split into sub--arrays with different targets;}
   \item{sky survey, with the aim of covering large portions of the sky to detect new or transient sources.}
\end{enumerate}
Case 1. applies mainly to flaring or varying sources, when the observation cannot be postponed and the maximum sensitivity over the whole energy range is required.
This is not the case of deep observations, since the comparison of the primary spectra with the expected sensitivity shows that the core energy window (around 1 TeV) will be covered in a time considerably smaller than the high energy tail.
We therefore expect for deep field observations that the SST array will make very long exposures on selected candidates suggested by MSTs on the basis of hard spectrum or high flux.
This is case 2., with the array split into the different telescope types, but mainly for MSTs, this can apply also to telescopes of the same type.
Case 3. is the most relevant one for this study, since it allows the detection of serendipitous sources, variable or unexpected, and GRBs are both.
This observation mode will be therefore analyzed in more details.

The current field of view (FoV) of operating 
IACT arrays (4--5$^{\circ}$) will be widened 
by MSTs and SSTs up to 8--10$^{\circ}$.
Unfortunately, since the detection of 
primary photons is through the Cherenkov 
images, the sensitivity is not uniform inside the FoV 
and dramatically drops towards its edge.
This means that, if we want to enlarge 
the total FoV by pointing the telescopes 
towards different directions,
the FoV of each telescope must overlap
in order to obtain a uniform sensitivity.
At present, only the Galactic plane 
was scanned, starting from HEGRA \cite{AHA01,AHA02}
up to the 1400 hours--long survey by H.E.S.S. \cite{AHA05,AHA06,REN09}.
Outside the Galactic plane, only small 
promising regions were observed, 
limiting our knowledge of the VHE sky to a few per cent.

\begin{figure}[t!]
\centering
\includegraphics[height=0.4\textheight,bb=0 0 238 350,clip]{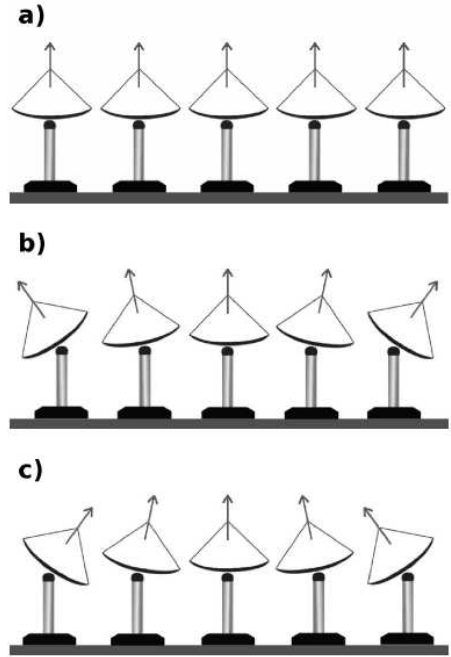}
\caption{Three modes of configuration of the telescope system used in the sky--survey scans: a) normal (parallel) mode; b) divergent mode; c) convergent mode (taken from \cite{divergent}).}
\label{fig:divergent}
\end{figure}

In order to efficiently scan large portions 
of the sky, the divergent mode was 
firstly proposed.
In this mode, instead of having all 
telescopes pointing towards the same direction 
(the so called ``parallel mode''), each telescope 
points to an angle slightly increasing 
from the center to the edges 
of the array.
Another possibility, the so called 
``convergent mode'', envisages that this 
angle is reduced from the array 
center to its edges (see the sketch in 
Figure \ref{fig:divergent}).
Both modes were deeply studied for MSTs 
by means of an accurate simulation 
of the shower development in the atmosphere 
and of the telescopes response 
in \cite{divergent}, for different 
angle separations.
The results show that the maximum 
sensitivity for sky survey is achieved 
in divergent mode with large 
offset angles ($\sim$ 6$^{\circ}$ from 
the central to the outermost telescopes), 
with a decrease of the observing time 
for a given sensitivity by a factor of 
$\sim$2.3 with respect to the parallel mode.
However, the angular and energy resolutions 
are worsened by a factor up to 2.
The convergent mode is a better 
choice at high energy, favoring 
the observation of sources with hard spectra.
In this work, we propose a third possibility, 
that is a mixture between parallel and divergent modes.

In the past, exciting observations made by Whipple 
in the late '80s were performed using a single 
telescope recording the Cherenkov images by 
means of a pixelated camera that allowed the 
discrimination between gammas and hadrons.
With just one image, the core location (and thus 
the primary energy) could be badly determined, 
limiting the detection inside the 
Cherenkov ``pool'' (with radius r$_p$ $\sim$ 120 m) where the 
lateral distribution of photons is approximately flat.
After imaging, the next fundamental improvement 
to the IACT technique was the stereo 
approach firstly used by HEGRA: with 
at least two images, the shower axis 
could be determined geometrically with a better 
resolution on both energy and arrival direction.
This approach was so widely adopted that even 
for CTA the main trigger will be given by 
the coincidence of at least two telescopes, 
and the mono events will be collected only 
for calibration and testing purposes.

Since for IACTs the single observing unit
is a couple of nearby telescopes, we propose here a 
``coupled divergent mode'' in which couples of 
telescopes are pointing to slightly different 
positions as in Figure \ref{fig:divergent}b) 
for single ones.
The tilt angle will be chosen on the basis of 
the sensitivity decrease over the FoV for each 
couple of telescopes.
If, for example, the sensitivity is 
halved at 3$^{\circ}$, a tilt angle of 
6$^{\circ}$ will assure a uniform sensitivity 
at least along the line connecting the center 
of the FoV of the two couples of telescopes.
For the entire sky we expect that this value 
will be reduced, and we plan to perform 
an accurate simulation 
to optimize this separation angle.
Even if the main contribution in GRB searches
will come from LSTs, which due to their paucity  
can hardly benefit of 
whatever divergent mode, the contribution of 
MSTs could still be not negligible.
The medium sized telescopes are not designed 
for fast slewing as LSTs (which can repoint within 
100 s or less), but there are 
indications from EGRET and from Fermi observations
that the VHE emission is delayed 
with respect to the prompt phase.
Concerning the energy window, the energy range 
from 100 GeV to 1 TeV corresponds respectively to an 
horizon from z $\simeq$ 1 to z $\simeq$ 0.1, 
depending on the EBL absorption model.
For follow--up observations, the GRB location 
is often given by fast satellite analysis with 
large uncertainties, mostly greater than the 
50\% sensitivity FoV, so some kind 
of divergent mode must be applied.
Moreover, a sky survey with large angular 
acceptance could reveal a serendipitous GRB.

In the future, Monte Carlo simulations
to obtain the expected 
performance of the MST array operated in 
``coupled divergent mode'' are planned, and the results 
will be compared with those presented in \cite{divergent}.
An estimate of the rate of serendipitous GRBs 
detectable during the sky survey will be also given.
\section{Acknowledgments}
We gratefully acknowledge support from the agencies 
and organizations 
listed under Funding Agencies at this website: 
{\tt http://www.cta-observatory.org/}.
\end{document}